\documentstyle[seceq,preprint,amssymb]{ptptex}

\def\be{\begin{equation}}
\def\ee{\end{equation}}
\def\bea{\begin{eqnarray}}
\def\eea{\end{eqnarray}}

\def\eq#1{(\ref{#1})}
\def\pt{\partial}
\newcommand{\CR}{\nonumber \\}

\preprintnumber[3cm]{
KUNS-1566\\
hep-th/9903146 }

\title{
An Uncertainty Relation of Space-Time
}

\author{
Naoki {\sc  Sasakura}\footnote{E-mail: sasakura@gauge.scphys.kyoto-u.ac.jp} 
}

\inst{
Department of Physics, Kyoto University, Kyoto 606-8502, Japan
}

\recdate{
}

\abst{
We propose an uncertainty relation of space-time. This relation is 
characterized by $G\hbar T \lesssim \delta V$, where $T$ 
and $\delta V$ denote
a characteristic time scale and a spatial volume, respectively.
Using this uncertainty relation, we give qualitative estimations for
the entropies of a black hole and our universe. We obtain qualitative
agreements with the known results.
The holographic principle of 't Hooft and Susskind is reproduced. 
We also discuss cosmology and give a relation to the 
cosmic holographic principle of Fischler and Susskind.
However, as for the maximal entropy of a system with an energy $E$, 
we obtain the formula $\sqrt{EV/G\hbar^2}$, with $V$ denoting the volume
of the system, 
which is distinct from the Bekenstein entropy formula $ER/\hbar$ with
$R$ denoting the length scale of the system.
}

\begin{document}

\maketitle

\section{Introduction}
\label{sec:intro}

Despite various noteworthy attempts,
the cosmological constant problem remains as one of the major 
mysteries in physics. \cite{wein} 
If we believe the locality of our world, the fundamental degrees of 
freedom of a system should be the degrees of freedom at a sufficiently 
small scale, because, understanding these, 
we can make a prediction at any scale larger than this small scale.
The natural small scale associated with gravity is the Planck length. 
The fundamental degrees of freedom of gravity are believed to be somehow 
associated with this Planck length. 
Our general knowledge of quantum field theory tells us that the vacuum
quantum fluctuations of these degrees of freedom induce a
cosmological constant on the huge order 
(Planck length)$^{-4}$, since a cosmological constant can exist in
general relativity, which is well established as the theory of
gravity for macroscopic phenomena.

On the other hand, turning to the matter sector, 
QCD is well established as a quantum field theory of $SU(3)$
non-abelian gauge fields (and quark fields).
QCD is an asymptotically free field theory, and  
non-abelian gauge fields are weakly interacting at a sufficiently
small scale.
On such a small scale, the non-abelian gauge field behaves like a
classical field, and we may safely assume that the fundamental degrees
of freedom of QCD are the (classical) non-abelian gauge field.
However, in quantum gravity, we have a conflict between the 
locality and the weakness of the interactions. 
Although the metric tensor field
describes macroscopic phenomena quite well,  
the interaction becomes stronger on a smaller scale.
If we impose the locality of our world, the fundamental degrees of
freedom of gravity should have quantum mechanical properties as their 
basic properties.
The quantum mechanical properties may be incorporated by an uncertainty 
relation of space-time.

In this paper, we propose a space-time uncertainty relation
and question whether the fundamental scale associated with gravity 
is really the Planck length.
Our proposal looks quite unusual at first sight, 
but it turns out to be consistent with known results.  
In \S\ref{sec:notion}, we explain the motivation for
our proposal.
Based on the space-time uncertainty relation, we find that the
density of the degrees of freedom vanishes in an infinitely stretching 
space-time. This
might be a nice property to solve the cosmological constant problem.
In \S\ref{sec:region}, we evaluate the entropy in a
spatial region as a function of the volume and the energy of the
region and compare it with the previously obtained results.
In \S\ref{sec:thermodynamics}, we discuss thermodynamics.
In \S\ref{sec:cosmology}, we evaluate the entropy and the
energy density originating from the space-time itself in our universe.
In \S\ref{sec:summary}, we summarize our results and give some
discussion.
In the Appendix, we give the derivation of the entropy formula used 
in \S\ref{sec:region}.
Throughout this paper, we use a unit system in which $c=1$.

\section{A space-time uncertainty relation}
\label{sec:notion}

The fundamental degrees of freedom of general relativity are the 
metric tensor field, which is associated with each point in the 
(3+1)-dimensional space-time.
It is implicitly assumed that we can construct a coordinate system and 
measure the values of the metric tensor field. 
This process seems to have no difficulties 
in classical mechanics. We could arrange ``clocks'' in an
appropriate way and measure the metric field by interchanging 
light or some particles among them.
As one arranges more ``clocks'' in a spatial region, the
measurement becomes finer. 
One might worry about the influence from the masses of 
the ``clocks'' themselves when the density of the ``clocks'' becomes
large.
But it would not contradict the principles of classical mechanics 
to make a ``clock'' with an arbitrarily small mass, though it would
become an unlimitedly difficult engineering problem.

However, in quantum mechanics, one notices a serious difficulty in
the above process.\cite{simtho} As the mass of each ``clock'' becomes 
smaller, the coordinate system decays in a shorter time. To see this, 
let us consider the construction of a coordinate system for a time
interval $T$ and with a spatial fineness $\delta x$ in a Minkowski space-time.
Since a ``clock'' must be localized in a region with the scale 
$\delta x$, the ``clock'' inevitably has a momentum of the order
$p\sim\hbar/\delta x$, obtained from the uncertainty relation of
quantum mechanics.
Thus the ``clocks'' move with a finite velocity of order 
$v\sim\hbar/m\delta x$, where $m$ denotes the masses of the ``clocks''.
This implies that the coordinate system will be destroyed by the quantum
effect in a finite
period $\delta x/v\sim m(\delta x)^2 / \hbar$.
This period must be larger than the time interval $T$ of the
coordinate. Hence we obtain 
\be
T\lesssim m(\delta x)^2/ \hbar.
\label{eq:ineqtx}
\ee
This gives a lower bound for the ``clock'' mass $m$ for given $T$ and 
$\delta x$.

{}From \eq{eq:ineqtx}, we  need ``clocks'' with a larger mass to construct a
finer coordinate system. However we also have a maximum value of a
``clock'' mass, because no ``clock'' should become a black
hole. To measure the gravitational field, it should be
possible to interchange light or particles among ``clocks''.
Thus the Schwarzschild radius should not exceed the uncertainty
$\delta x$ of each ``clock'':
\be
Gm \lesssim \delta x.
\label{eq:gmx}
\ee
The ``clock'' mass can be chosen arbitrary if it satisfies
\eq{eq:ineqtx} and \eq{eq:gmx}.
Thus the condition for the existence of such a ``clock'' mass is 
\be
G\hbar T \lesssim (\delta x)^3.
\label{eq:txxx}
\ee

Since we cannot construct an appropriate coordinate system unless 
the inequality \eq{eq:txxx} is satisfied, it would be natural to propose that 
the inequality \eq{eq:txxx} is the property of a space-time
itself with $T$ and $\delta x$ denoting the time scale and the spatial
uncertainty of the space-time, respectively.
Regarding $(\delta x)^3$ as the volume associated with each ``clock'', 
\eq{eq:txxx} would be rewritten in the form   
\be
G\hbar T \lesssim \delta V,
\label{eq:tv}
\ee
where $\delta V$ denotes a volume uncertainty.

A flaw in the above derivation is that we neglected the dynamical
effect caused by the gravitational field generated by the ``clocks''.
One easily finds from \eq{eq:ineqtx} that a coordinate system is 
destroyed in a much shorter time by the 
gravitational effects among the ``clocks'' themselves than by the quantum
effects. To justify neglecting this effect, we should consider the
case in which the 
distances among ``clocks'' are much larger than the fineness $\delta x$.
In this case, the fineness $\delta x$ represents the uncertainty of the 
coordinate associated to each ``clock''. 

An unusual point of the inequality \eq{eq:txxx} is that the minimal spatial
length $\delta x$ has a positive correlation with the time interval $T$.
Although it seems as if the left-hand side of \eq{eq:txxx} can be made
arbitrarily small, we cannot detect an arbitrarily small spatial
length scale. To show this, let us suppose a test particle is injected 
into a system to detect a structure with a length scale $\delta x$ 
of the system by a collision experiment.
Since the speed of the particle is 
smaller than the speed of light, the collision 
experiment needs a time longer than $\delta x$. Thus, from \eq{eq:txxx},
we have
\be
G\hbar \delta x < G\hbar T \lesssim (\delta x)^3,
\ee
which dictates that the minimal length detectable is indeed the Planck
length $\sqrt{G\hbar}$.

When we take the time scale $T$ very large, the inequality
\eq{eq:txxx} tells us quite an unusual thing for a space-time.
If the time scale $T$ diverges, the fineness $\delta x$ also diverges.
This means that a space-time with an infinite characteristic time
scale has a vanishing density of degrees of freedom.
We will use this property in the discussion of the
cosmological constant problem in \S\ref{sec:cosmology}.

{}From \eq{eq:tv}, it would be natural to propose that a fundamental
degree of freedom of quantum gravity is associated with a finite spatial
volume determined by 
\be
\delta V = c_{bt} G\hbar T,
\label{eq:bit}
\ee
where $c_{bt}$ is a numerical constant. We discuss the value of 
this constant in \S\ref{sec:cosmology}, but for the time being 
we simply leave it undetermined.
We call each such fundamental space-time region a ``space-time bit''
in this paper.

\section{The number of bits in a region with a given energy}
\label{sec:region}

In this section, we discuss the maximal number of the space-time bits
in a spatial region with a total volume $V$ and a total energy $E$, and
give a formula for the entropy associated with the region.

Suppose a spatial region is composed of bits labeled by an integer
$i=1,\cdots,N$. We assume the total spatial volume $V$ is partitioned 
by the bits
\be
V=\sum_{i=1}^N \delta V_i.
\label{eq:v}
\ee
We now use the uncertainty inequality for energy, $\delta E \geq \hbar / T$.
We assume also that the total energy is partitioned by the bits.
Then we have
\be
E=\sum_{i=1}^N \delta E_i \gtrsim \sum_{i=1}^N \frac\hbar{T_i}
\sim \sum_{i=1}^N \frac{G \hbar^2}{\delta V_i}.
\label{eq:e}
\ee
In the last relation, we have used \eq{eq:bit}.  
Under the constraint \eq{eq:v},
the quantity  $\sum_{i=1}^N \frac{G \hbar^2}{\delta V_i}$ in \eq{eq:e} 
takes its minimum value when the volume
of each bit, $\delta V_i$, takes the same value $V/N$.
Thus we obtain an inequality
\be
E\gtrsim \frac{G\hbar^2N^2}{V},
\ee
or
\be
N\lesssim \sqrt{\frac{EV}{G\hbar^2}}.
\label{eq:EV}
\ee
There is no reason to believe that 
the maximum value of the number of the bits in a region
agrees with the entropy associated with it. However, in the Appendix, 
we give discussion that illustrates that the right-hand side of
\eq{eq:EV} is in fact approximately proportional to the
entropy:\footnote{The entropy in this paper is dimensionless.}
\be
S(E,V)\sim c_0 \sqrt{\frac{EV}{G\hbar^2}},
\label{eq:ent}
\ee
where $c_0$ is a proportionality factor that is a constant in the 
leading approximation. 

The right-hand side of \eq{eq:EV} becomes $\sqrt{ER^3/G\hbar^2}$
for a system with a characteristic scale $R$.
Thus the equation \eq{eq:EV} has an extra factor $\sqrt{R/EG}$,
compared with the so-called Bekenstein entropy bound 
$ER/\hbar$.\cite{bek,revbek}
In the case that the Schwarzschild bound $GE<R$ is satisfied, 
the bound \eq{eq:EV} is larger than
the Bekenstein entropy bound.
Thus \eq{eq:EV} is distinct from the Bekenstein entropy bound but
does not contradict it. In the case $GE>R$, the entropy formula
\eq{eq:ent} gives a value smaller than the Bekenstein bound.
This plays an important role in explaining the entropy of our universe
in \S\ref{sec:cosmology}. 
  
Substituting \eq{eq:EV} with the Schwarzschild bound $GE<R$, we obtain 
\be
N\lesssim \frac{R^2}{G\hbar}.
\ee
This is the basic inequality of the holographic principle
of 't Hooft and Susskind \cite{hol} that the world can be described by 
the degrees of freedom on a two-dimensional surface.    
In the case of a black hole, the Schwarzschild bound is saturated.
Regarding $R^2$ as the area of the black hole, we obtain qualitative 
agreement with the Bekenstein-Hawking entropy formula.\cite{rad,revbek}

\section{Thermodynamics}
\label{sec:thermodynamics}

In this section we discuss thermodynamics based on the
entropy formula \eq{eq:ent}.

Using the first law of the thermodynamics $dE=TdS-PdV$, we find
\be
T=\left(\frac{\pt S}{\pt E}\right)^{-1}
=\frac{2}{c_0}\sqrt{\frac{G\hbar^2E}{V}},
\ee
and
\be
P=T\left(\frac{\pt S}{\pt V}\right)=\frac{E}{V}.
\label{eq:p}
\ee
If we assume that the energy distribution is uniform, from \eq{eq:p}
we obtain 
\be
P=\rho,
\label{eq:prho}
\ee
where $\rho$ denotes the energy density.  
If we regard $R$ as $V^{1/3}$,  we instead obtain
$P=E/3V=\rho/3$ from the Bekenstein entropy formula 
$S \sim ER$. 
This is the same relation between 
the energy density and the pressure as that of radiation, while we obtain
a distinct relation \eq{eq:prho}. The relation \eq{eq:prho} describes
the most incompressible fluid that is consistent with special
relativity. In this fluid, ``sound'' 
propagates at the velocity of light. 
This light speed propagation might be understood as that of a graviton.
Equation \eq{eq:prho} is distinct from the relation 
$P=-\rho=-\Lambda$, also, which we
would have if the cosmological constant term were interpreted as 
an energy-momentum tensor. 

The pressure \eq{eq:prho} should also play important roles in black
hole physics and cosmology.

\section{Cosmological implications}
\label{sec:cosmology}

In this section, we discuss the entropy of our universe
and then discuss the cosmological constant problem.

The total entropy of our universe would be accurately estimated by the entropy 
produced at the Planckian time $t_P=l_P=\sqrt{G\hbar}$ from the big bang.
Substituting $T=t_P$ into \eq{eq:bit}, the volume $\delta V_P$ of a
bit at the Planckian time is 
\be
\delta V_P \sim G\hbar t_P=(l_P)^3.
\ee
Thus the total number of the bits in the universe at the Planckian
time is estimated as 
\be
N_P \sim \frac{R_P^3}{\delta V_P} \sim \left(\frac{R_P}{l_P}\right)^3, 
\ee
where $R_P$ denotes the size of the observable part of our
universe at the Planckian time.
Thus, using \eq{eq:ent} and $E\sim N_P \hbar /t_P$, 
we obtain the entropy of our universe as 
\be
S_U\sim \sqrt{\frac{N_P V}{t_p G \hbar}}\sim N_P \sim 
\left(\frac{R_P}{l_P}\right)^3\sim 10^{90}.
\label{eq:entuni}
\ee
Here we have used $R_P\sim 10^{30} l_P$ obtained from the 
Friedman-Robertson-Walker (FRW) cosmological model with the
epoch of matter-radiation density equality $t_{\rm eq}\sim 10^{11}$ sec. 
The value \eq{eq:entuni} is very near the entropy obtained from the 
2.7 K cosmic microwave background.

This value \eq{eq:entuni} of the entropy of our universe has 
been recently discussed from the point of 
view of the cosmic holographic principle,\cite{coshol,cosent,cossec} 
Fischler and Susskind \cite{coshol} have shown that the cosmic
holographic principle gives a bound on the expansion rate and 
that it can be translated into a bound on the equation of state.
The bound is saturated by the most incompressible perfect fluid,
which we discussed in \S\ref{sec:thermodynamics}.  
We can obtain the same bound by imposing the second law of
thermodynamics, namely that the entropy of our universe should not decrease
with time. To show this, let us assume $a(t)\sim t^p$, where
$a(t)$ is the scale factor of the FRW metric. Then, $\delta V(t) \sim
t$ from \eq{eq:bit} and $V(t)\sim t^{3p}$. 
Thus, by a similar argument as that used in deriving 
\eq{eq:entuni}, we find the behavior of entropy to be described by  
\be
S(t)\sim N(t) \sim  t^{3p-1}.
\label{eq:entbeh}
\ee
Thus, for the second law to be satisfied,  we must have 
\be
p\geq\frac13,
\ee
which can be translated into the bound $\gamma \leq 1$ for 
the equation of state 
$P=\gamma \rho$. Relations between the generalized second law and 
the cosmic holographic principle are discussed in Ref.~\citen{cossec}.

As discussed in \S\ref{sec:notion}, the notion of a space-time
bit seems to seriously contradict the classical notion of 
the (3+1)-dimensional space-time, if the range of time is infinite.
This suggests that the space-time volume ``operator''
$\int d^4x \sqrt{-g}$ does not exist in ``quantum gravity''.  
Since the cosmological constant is the coupling constant associated with
this ``operator'', the suggestion indicates that we cannot introduce 
the cosmological constant into ``quantum gravity''.
However, our universe does not have an infinite range of time, 
so we expect there to be a term analogous to the cosmological constant. 
We can estimate this as follows.
We choose the age of the universe $T_U$ as the 
characteristic time scale.
Then, from \eq{eq:bit}, the volume of a space-time bit of our universe is 
\be
\delta V_U\sim c_{bt}G\hbar T_U.
\ee
Each bit will have an energy $\hbar/T_U$, from the uncertainty
relation.
Thus the energy density of our universe originating from the space-time
itself is obtained as 
\be
\rho_{st}\sim\frac{\hbar}{T_U \delta V_U}\sim\frac{1}{c_{bt}GT_U^2}.
\label{eq:coscon}
\ee
This is on the same order as the critical density 
$\rho_c=3H^2/8\pi G$, where $H$ is the Hubble constant, and so 
might be large enough to change significantly the evolution scenario of
our universe.

We now give an argument to estimate the numerical constant $c_{bt}$ in
\eq{eq:bit}.
Let us consider an FRW cosmological model with a vanishing
spatial curvature, and, as a matter, consider a perfect fluid
with the thermodynamic property given in \S\ref{sec:thermodynamics}.
The conservation law of the stress-energy tensor, 
$\dot\rho+3(\rho+P)\dot a/a =0$, determines the density as 
\be
\rho=\frac{C}{a^6},
\label{eq:conrho}
\ee
where $C$ is a numerical constant.
Substituting \eq{eq:conrho} into the evolution equation
$3\dot a ^2/a^2=8\pi G\rho$, we obtain $a^3=\sqrt{24\pi G C}t$, and hence
\be
\rho=\frac{1}{24\pi G  t^2}.
\label{eq:rhoein}
\ee
Since general relativity gives a perfect description, at least 
for a macroscopic object,
it would be natural to demand that the result \eq{eq:rhoein}  
agree with \eq{eq:coscon} for $T_U=t$. Moreover, since 
$a(t)\sim t^{1/3}$, this evolution is an adiabatic process, as can be
seen from \eq{eq:entbeh}, and is consistent with the conservation of 
the stress-energy tensor used in the derivation of \eq{eq:conrho}.
Thus we obtain
\be
c_{bt}\sim 24\pi.
\label{eq:cbt}
\ee

An important comment is that the perfect fluid with the thermodynamic
properties in section \ref{sec:thermodynamics} does not fully 
represent the properties of space-time bits. As an example in the 
FRW model of a flat spatial curvature, 
let us consider the case in which there exists dust.
Since the energy density of dust behaves as $1/a^3$ and decreases more
slowly than \eq{eq:conrho}, the evolution of the universe is
dominated by dust after a sufficiently long time. Then the scale factor 
behaves as $a\sim t^{2/3}$ in this regime, and, substituting this into 
\eq{eq:conrho}, we obtain $\rho\sim 1/t^4$. This contradicts 
\eq{eq:coscon}, which was derived simply from the uncertainty relation
of energy and time. We propose that the thermodynamic properties
change somehow in the regime and \eq{eq:coscon} is the correct answer.
Substituting \eq{eq:cbt} into \eq{eq:coscon},  
the ratio of the energy density originating from the space-time itself to the
critical density becomes
\be
\Omega_{st}=\frac{\rho_{st}}{\rho_c}\sim\frac{1}{9(HT_U)^2}.
\ee
Comparing this with the observational data, since $HT_U\sim 1$,  
we see that a good portion of the total energy density of 
our universe originates from the space-time itself.

Since the time scale of our universe is very large, it would be
interesting to estimate the length scale of a space-time bit.
We obtain
\be
(c_{bt}G\hbar T_U)^{1/3}\sim 10^{-14} {\rm\ m} 
\sim (10 {\rm\ MeV})^{-1},
\ee
where we have taken $T_U=10^{10}$ year.
Although this energy scale is much lower than the Planck energy
and is within the range of high energy experiments,
it is not clear whether or how this length scale can be observed in a
high energy experiment or an astrophysical observation.

\section{Summary and discussion}
\label{sec:summary}

In this paper, we have proposed an uncertainty relation for space-time
and have estimated some quantities in quantum gravity. An interesting 
point is that, although the proposed uncertainty relation is very  
different from the
usually expected relations such as $\delta t \delta x \gtrsim l_P^2$
or $\delta x \gtrsim l_P$, 
we have obtained results qualitatively consistent 
with the known results, and have not found any serious deviations from them. 
The purpose of this paper is merely to show the surprise of the
consistencies, but not 
to give a concrete way to calculate quantities in quantum gravity. 

Of course we hope that the uncertainty relation we have proposed 
turns out to be an intermediate notion which catches an essential property 
of quantum gravity. The most peculiar point is that our space-time
uncertainty relation is not consistent with the cosmological 
constant term. Presently the most promising approach to quantum gravity 
is string theory. 
The microscopic derivation of the entropy formula for a BPS black hole
{}from D-brane dynamics is impressive.\cite{entstr} 
The uncertainty relations in string theory are of the form 
$\delta t \delta x \gtrsim l_s^2$ and $\delta x \gtrsim l_{\rm min}$,\cite{yon}
which work evidently as an ultraviolet cutoff.
The reduction of the degrees of freedom in the infrared limit is 
realized through a duality which interchanges small scale and 
large scale dynamics.\cite{suswit,yon}
Thus there is the possibility for an uncertainty relation similar
to ours to be derived from string theory.
We hope this as well as our space-time uncertainty relation will 
be helpful in constructing a new theory describing quantum gravity.

Finally, in lattice approaches to quantum gravity, an uncertainty 
relation of the kind $\delta x,\delta t \gtrsim l_{\rm min}$ 
seems to be assumed implicitly. \cite{latgra}
Our uncertainty relation may provide a new direction for such approaches.

\section*{Acknowledgements}
We would like to thank H.~Hata for carefully reading the manuscript.
This work was supported in part by a Grant-in-Aid for Scientific Research
{}from Ministry of Education, Science, Sports and Culture (\#09640346) and
Priority Area: ``Supersymmetry and Unified Theory of Elementary
Particles'' (\#707). 

\appendix
\section{} 
In this appendix, we evaluate a quantity which is similar to 
entropy in statistical mechanics to illustrate that the right-hand side
of \eq{eq:EV} can be regarded as a quantity proportional to the entropy
of space-time. What we consider here is just an example. It is not a
proof, because we do not have any reliable microscopic theory of
quantum gravity. 

Let us consider a space-time with a total number of bits $N$, a total 
volume $V$ and a total energy $E$.
The quantity we evaluate is the phase volume of energies and volumes with  
the constraint \eq{eq:tv}:  
\be
\Omega(N,V,E)=\prod_{i=1}^N \left(\int_0^\infty\int_0^\infty 
\frac{dV_idE_i}{G\hbar^2}
\theta\left(V_iE_i-G\hbar^2\right)\right) 
\delta\left(\sum_{i=1}^N V_i-V\right)
\delta\left(\sum_{i=1}^N E_i-E\right),
\label{eq:onve}
\ee
where $\theta(y)$ denotes the step function: $\theta(y)=1$ for $y\geq 0$
and otherwise vanishing.
The Laplace transform of \eq{eq:onve} is evaluated as 
\bea
\Omega(N,\alpha,\beta)&=& \int_0^\infty\int_0^\infty
\frac{dVdE}{G\hbar^2} e^{-\alpha V - \beta E} \Omega(N,V,E) \CR
&=& \prod_{i=1}^N \int_0^\infty\int_0^\infty \frac{dV_idE_i}{G\hbar^2}
e^{-\alpha V_i-\beta E_i} \theta\left(V_iE_i-G\hbar^2\right) \CR
&=& \left[ \int_0^\infty 
\frac{dV}{G\hbar^2 \beta} e^{-G\hbar^2\beta/V-\alpha V}\right]^N \CR
&=&\left[\frac2{\sqrt{G\hbar^2\alpha\beta}}
K_1(2\sqrt{G\hbar^2\alpha\beta})\right]^N \CR
&=& \left[\frac{\sqrt{\pi}}{(G\hbar^2\alpha\beta)^{3/4}}
\exp\left(-2\sqrt{G\hbar^2\alpha\beta}\right)
h\left(2\sqrt{G\hbar^2\alpha\beta}\right)\right]^N,
\label{eq:lapfin}
\eea
where $K_1(z)$ is a Bessel function of imaginary argument, and 
$h(z)$ is a function with the following asymptotic series 
for large $z$:
\be
h(z)=\sum_{n=0}^{\infty} \frac{\Gamma(3/2+n)}{n!\Gamma(3/2-n)(2z)^n}.
\ee
The function $h(z)$ can be approximated as $1$ if $z$ is sufficiently large.
In the calculation below, we neglect $h(z)$. The consistency
of this simplification is checked later.

The quantity $\Omega(N,V,E)$ is  obtained by the inverse Laplace
transform of \eq{eq:lapfin}:
\be
\Omega(N,V,E)=
\int_{-i\infty}^{i\infty}\int_{-i\infty}^{i\infty} d\alpha d\beta\ 
\Omega(N,\alpha,\beta)\exp\left(\alpha V + \beta E\right).
\ee
The integration can be approximated by the saddle point method.
The saddle point is determined by
\bea
S(N,\alpha,\beta) &\equiv& N\left(\frac12 \ln(\pi) - \frac34
\ln(G\hbar^2\alpha\beta) -2\sqrt{G\hbar^2\alpha\beta}\right)
+\alpha V + \beta E, \CR
\frac{\pt S(N,\alpha,\beta)}{\pt \alpha}&=&
\frac{\pt S(N,\alpha,\beta)}{\pt \beta}=0.
\eea
The solution is 
\bea
\alpha_0&=& \frac{3N}{4V(1-\sqrt{G\hbar^2N^2/VE})}, \CR
\beta_0&=& \frac{3N}{4E(1-\sqrt{G\hbar^2N^2/VE})}. 
\eea
Substituting this back, we obtain
\bea
\ln(\Omega(N,V,E))&\sim& S(N,\alpha_0,\beta_0) 
=N_0f\left(\frac{N}{N_0}\right), 
\label{eq:entnapp}\\
f(z)&=&\left(\frac12 \ln(\pi)-\frac32 \ln\left(\frac34\right)+\frac32
\right)z
-\frac32z \ln\left(\frac{z}{1-z}\right), \CR
N_0&=&\sqrt{\frac{EV}{G\hbar^2}}.
\eea

For given $E$ and $V$,
the maximum value of \eq{eq:entnapp} with respect to $N$ will give 
its total entropy. The function $f(z)$ takes its maximum 
value $f(z_0)\sim 1.3$ at $z_0\sim 0.5$. The error introduced by neglecting
$h(z)$ is just $z_0\ln(h(z_0))\sim 0.1$, and hence is of next higher order. 
Thus we obtain
\be
S(V,E)\sim 1.4 \sqrt{\frac{EV}{G\hbar^2}}.
\ee

\vspace{.5cm}
\noindent
Note added:
Thought experiments similar to that in \S\ref{sec:notion}
appeared in Refs.~\citen{simtho}
to derive the uncertainty in the measurement of a space-time distance.
We would like the thank Y.\ J.\ Ng for this information.
We would also like to thank T.\ Yoneya for explaining in detail the 
stringy uncertainty relation and informing us of some references. 

\end{document}